\documentclass[%
 aip,
 amsmath,amssymb,
 reprint,%
]{revtex4-1}

\usepackage{graphicx}
\usepackage{dcolumn}
\usepackage{bm}
\usepackage{float}

\usepackage[utf8]{inputenc}
\usepackage[T1]{fontenc}
\usepackage{mathptmx}
\usepackage{etoolbox}
\usepackage[normalem]{ulem}

\usepackage{xcolor} 

\newcommand{\ks}{\textcolor{black}} 

\def\We{{\it We}}
\def\d{{\rm d}}

\def\R{{\cal R}}

\makeatletter
\def\@email#1#2{%
 \endgroup
 \patchcmd{\titleblock@produce}
  {\frontmatter@RRAPformat}
  {\frontmatter@RRAPformat{\produce@RRAP{*#1\href{mailto:#2}{#2}}}\frontmatter@RRAPformat}
  {}{}
}%
\makeatother
\begin{document}

\preprint{AIP/123-QED}

\title{Drop size distribution from laboratory experiments based on single-drop fragmentation and comparison with aerial in-situ measurements}
\author{Shubham Chakraborty}
 \affiliation{Center for Interdisciplinary Program, Indian Institute of Technology Hyderabad, Telangana - 502 284, India}
 \author{Someshwar Sanjay Ade}
 \affiliation{Center for Interdisciplinary Program, Indian Institute of Technology Hyderabad, Telangana - 502 284, India}
\author{Lalsingh Devsoth}
 \affiliation{Department of Mechanical and Aerospace Engineering, Indian Institute of Technology Hyderabad, Telangana - 502 284, India}
 \author{Lakshmana D. Chandrala}
 \affiliation{Department of Mechanical and Aerospace Engineering, Indian Institute of Technology Hyderabad, Telangana - 502 284, India}
 \author{Thara Prabhakaran}
 \affiliation{Indian Institute of Tropical Meteorology (Ministry of Earth Sciences, New Delhi), Pune, India}
  \author{Omar K. Matar}
 \affiliation{Department of Chemical Engineering, Imperial College London, London SW7 2AZ, UK}
 \author{Kirti Chandra Sahu}
 \affiliation{Department of Chemical Engineering, Indian Institute of Technology Hyderabad, Kandi - 502 284, Telangana, India}
\email{ksahu@che.iith.ac.in}

\date{\today}

\begin{abstract}
Laboratory experiments and theoretical modelling are conducted to determine the raindrop size distribution (DSD) resulting from distinct fragmentation processes under various upward airstreams. Since weather radar echoes are proportional to the sixth power of the average droplet diameter, understanding the fragmentation mechanisms that lead to different breakup sizes is crucial for accurate rainfall predictions. We utilize a two-parameter gamma distribution for theoretical modelling and estimate the average droplet diameter from the theoretically obtained characteristic sizes, often treated as assumed input parameters for different rain conditions in rainfall modelling. Our experimental and theoretical findings demonstrate a close agreement with the DSD predicted by the Marshall and Palmer relationship for steady rain conditions. Additionally, in situ DSD measurements at different altitudes were obtained through research flights equipped with advanced sensors, further validating our rainfall model. This study underscores the effectiveness of laboratory-scale experiments and the critical importance of accurately characterizing DSD to enhance rainfall predictions.
\end{abstract}

\maketitle

\section{Introduction}
\label{sec:intro}

\begin{figure}
\centering
\includegraphics[width=0.45\textwidth]{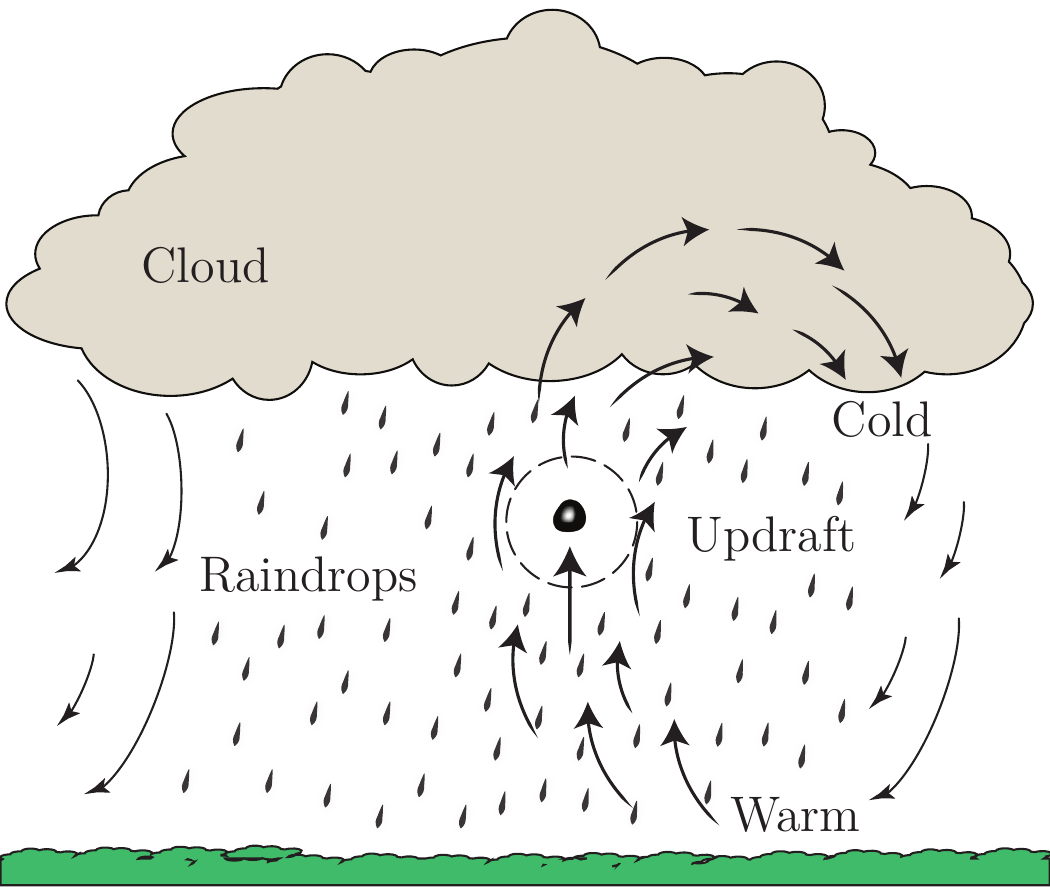}
\caption{Updraft (upward movement) of warm air from the surface during rain interacting with raindrops in a counter-current configuration.}
\label{clouds}
\end{figure}

Accurate rainfall prediction remains one of the major challenges in environmental research due to its critical role in understanding climate change and its wide-ranging socio-economic implications. The interaction between airstreams and raindrops during rainfall influences the shape and size distribution of raindrops, which are critical factors in rainfall forecasting and play a pivotal role in the development of heavy rainfall, thunderstorms, and hail. The interaction of the airstream with raindrops is characterized by the Weber number, a dimensionless parameter in fluid mechanics that compares the relative significance of fluid inertia to surface tension. The Weber number is given by $\We \equiv \rho_a U^2 d_0/\sigma$, where $\rho_a$ is the air density, $\sigma$ is the interfacial tension, $U$ is the average velocity of the airstream, and $d_0$ is the droplet diameter. In a typical rainfall, the upward air movement (updraft), as illustrated in Figure \ref{clouds}, generally ranges from 15 m/s to 20 m/s, though it can occasionally reach up to 50 m/s \citep{marinescu2020updraft,musil1986microphysical}. Thus, for raindrops with sizes between 2 mm and 8 mm, the Weber number ranges from 1 to 100, leading to distinct fragmentation processes.

The fragmentation of droplets produces multiple sizes of child drops, resulting in a droplet size distribution (DSD) within a specific air volume. The resultant DSD is crucial for determining rainfall intensity, interpreting radar data, and improving hydrological models \citep{uijlenhoet2001raindrop}. The radar echo from a raindrop is highly sensitive to its size, with reflectivity increasing by the sixth power of the diameter ($d^6$), meaning that larger drops have a significantly greater impact on radar reflectivity. Furthermore, DSD is essential for understanding physical processes like coalescence, breakup, and evaporation. \citet{laws1943relation} were the first to report the DSD under various rain conditions. Observations by \citet{marshall1948distribution} in a relatively steady rain revealed that the DSD, $n(d)$, follows an approximate negative-exponential form:
\begin{equation}
n(d) = n_0 e^{-d/d_{avg}}, \label{eq:mp}
\end{equation}
where $n_0$ is the average spatial density of drops (measured in cm$^{-4}$) and $d_{avg}^{-1} = \Lambda \R^{-a}$. Here, $\R$ represents the rainfall rate in mm/hr and $d_{avg}$ denotes the average diameter in centimetres (cm). The variables $\Lambda$ and $a$ are fitting parameters that characterize the relationship between raindrop size and rainfall intensity. In practice, scientists often assume the values of $n_0$, $\Lambda$, and $a$ for different rain conditions. \citet{marshall1948distribution} found that for steady rainfall, $n_0=0.08$ cm$^{-4}$, $\Lambda=41$, and $a=0.21$. Since then, various models have been developed to describe the DSD under different rainfall conditions based on the assumption of an average droplet diameter and other key parameters \citep{yau1996short, Villermaux2009single, mezhericher2022size}.

In the context of atomization, the fragmentation of droplets in a horizontal airstream has been extensively studied for several decades \citep{pilch1987use,cao2007new,guildenbecher2009secondary,suryaprakash2019secondary,soni2020deformation,kirar2022experimental}. These studies have demonstrated that at low Weber numbers, droplets exhibit shape oscillations (vibrational mode). As the Weber number increases, the droplet forms a single bag on the leeward side, surrounded by a thick liquid rim. The fragmentation of the bag and rim is driven by Rayleigh-Plateau (RP) capillary instability, Rayleigh-Taylor (RT) instability, and the nonlinear instability of liquid ligaments, leading to the formation of droplets of various sizes \citep{taylor1963shape,jackiw2021aerodynamic,jackiw2022prediction}. Specifically, the Rayleigh-Plateau instability breaks a liquid column into droplets as surface tension minimizes surface energy, while the Rayleigh-Taylor instability arises when a denser fluid accelerates into a lighter one, destabilizing the interface. The transition between the vibrational and bag breakup in a cross-stream configuration occurs at $\We \approx 12$. Variations of bag breakup include bag-stamen and multi-bag breakup occurring at higher Weber numbers. 

In shear mode, the droplet forms a downstream sheet that breaks into tiny child droplets. At high Weber numbers, droplets undergo catastrophic breakup, rapidly exploding into fragments. All these fragmentation processes lead to different droplet size distributions. Recently, \citet{ade2023size} have demonstrated that single-bag breakup produces a tri-modal distribution, while dual-bag breakup results in a bi-modal distribution. Droplet interaction with a counter-current airstream also results in similar morphologies, albeit at lower Weber numbers. For instance, in a counter-current airstream, the bag fragmentation of droplets occurs at $\We \approx 6$, as the droplet remains within the potential core region of the airstream \citep{Villermaux2009single}. This situation is analogous to a raindrop falling in a quiescent air medium \citep{agrawal2017nonspherical,balla2019shape}. It is worth noting that raindrops frequently form within cloud updrafts and later transition to downdraft regions near the cloud edges, where the typical Weber numbers, particularly for smaller raindrops, tend to be lower \citep{khain2013mechanism}. In the present study, we investigate the DSD resulting from various fragmentation processes by considering a broader range of Weber numbers ($\We > 12$), typically observed in convective systems.

The theory proposed by \citet{Villermaux2009single} suggests that single-drop bag fragmentation provides sufficient information to determine rain intensity. A single-parameter gamma size distribution, based solely on the capillary instability of the corrugated ligament, was employed to characterize the drop size distribution resulting from single-bag fragmentation. They derived a  relationship between average droplet diameter $(d_{avg})$ and rainfall rate $(\R)$, which is given by:
\begin{equation} 
\R = n_0 \frac{\pi}{6} \sqrt{\frac{\rho_w}{\rho_a}} \sqrt{g}{({d_{avg}})^\frac{9}{2}} \int x^{\frac{7}{2}} P(x) \, dx, \label{eq:davg_rainfall_rate} 
\end{equation} 
and $d_{avg}^{-1} = 48.5 \R^{-2/9}$ (i.e. $\Lambda=48.5$ and $a=2/9$). In Eq. (\ref{eq:davg_rainfall_rate}), $g$ denotes the acceleration due to gravity (9.81 m/s$^2$), and $\rho_w ~(=998$ {\rm kg/m}$^3)$ and $\rho_a ~(=1.2 ~{\rm kg/m}^3)$ represent the densities of water and air, respectively. In this single-parameter gamma distribution framework, the number size distribution, $P(x)$, of child droplets with normalized size $x = d/d_{avg}$ is expressed as:
\begin{equation} 
P(x) = \frac{32}{3}x^{3/2}K_{3}(4\sqrt{x}), \label{eq:villermaux_equation} 
\end{equation}
where $K_3$ is the modified Bessel function of the second kind \citep{Villermaux2009single}. This theory employed the average droplet diameter as an input to the rainfall model, which approximately correlates with the Marshall-Palmer formula (Eq. \ref{eq:mp}) derived for steady rain conditions. However, the model of \citet{Villermaux2009single} may not adequately estimate rainfall under varying atmospheric conditions. Moreover, the average diameter of raindrops, essential for measuring rainfall, is unknown a priori. Additionally, droplet fragmentation is not driven by a single instability; rather, it involves multiple distinct instabilities acting simultaneously \citep{jackiw2022prediction}, which further complicates the accurate estimation of rainfall intensity. For instance, the bag breakup process involves three distinct fragmentation modes: bag, receding rim, and node. Various instabilities, including the RT instability, capillary instability, receding rim instability, and the nonlinear instability of liquid ligaments, influence each of these breakup modes. These fragmentation mechanisms generate a range of characteristic breakup sizes that can affect rainfall calculations. Therefore, it is necessary to consider all these instabilities and their corresponding breakup sizes at different Weber numbers typically observed in atmospheric conditions during various rain conditions. 

\begin{figure*}
\centering
\includegraphics[width=0.7\textwidth]{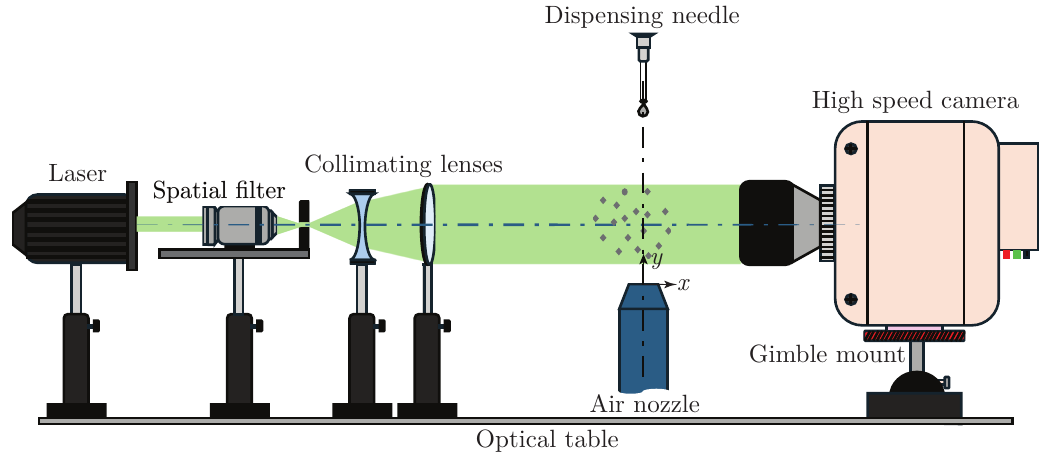}
\caption{Schematic diagram of the experimental setup (side view) used in in-line holography to obtain the drop size distribution.}
\label{holography_setup_schematic}
\end{figure*}

The present study addresses this critical gap in rainfall modelling by measuring the DSD and consequently estimating the average diameter of the distribution using the in-line holography technique supplemented by a theoretical model. More importantly, in situ aerial measurements of the raindrop size distribution were gathered through research flights equipped with advanced sensors, and the results were correlated with laboratory experiments to improve the understanding of DSD dynamics.

\section{Experimental method} \label{sec:method}

In the present study, a shadowgraphy technique is used to investigate the fragmentation of droplets subjected to vertical airstreams, characterized by the Weber number $(\We = \rho_a U^2 d_0/\sigma)$. Additionally, a machine-learning-based digital in-line holography technique is employed to analyze the size distribution of the resulting child droplets. The experimental setup consists of (i) an air nozzle (18 mm diameter), (ii) a droplet dispensing needle mounted on a three-dimensional (3D) traverse system, (iii) a continuous wave laser, (iv) a spatial filter assembly, (v) collimating optics comprising concave and convex lenses, (vi) two high-speed cameras, and (vii) a diffused backlit illumination source. A water drop of diameter $d_0 = 3.6 \pm 0.08$ mm is injected from the needle into an ascending airstream. The air nozzle (18 mm in diameter) is aligned with the needle along the same central axis, as shown in figure \ref{holography_setup_schematic}. The vertical airstream velocity is varied, with airflow controlled by a digital mass flow controller. The different airstream velocities considered in the present study are $U=12.44$ m/s, $U=16.7$ m/s and $U=17.68$ m/s, which correspond to $\We = 9.38$, 16.9 and 18.9, respectively. A high-speed camera (resolution: $2048 \times 1600$ pixels) with a Nikkor lens (focal length of 135 mm and minimum aperture of $f/2$) is employed for shadowgraphy. A high-power light-emitting diode (model: MultiLED QT, Make: GSVITEC, Germany) is used along with a uniform diffuser sheet to illuminate the background. The images are captured at 1800 frames per second (fps) with an exposure duration of 1 $\mu$s and a spatial resolution of 31.88 $\mu$m/pixel.

The digital in-line holography employs a collimated, coherent laser beam to capture interference patterns created by the scattered light from droplets (object wave) and the unscattered background light (reference wave) on a camera sensor. The recorded hologram contains both the amplitude and phase information of the object wave, which is then numerically reconstructed by simulating illumination with a reference beam, allowing droplet information to be retrieved at various depths. In this configuration, a single beam acts as both the reference and object wave. The digital in-line holography setup, as illustrated in figure \ref{holography_setup_schematic}, consists of a continuous wave laser, a spatial filter system, collimating optics, and a high-speed camera. The laser emits a 532 nm wavelength beam, which passes through a spatial filter to reduce noise. The beam is then expanded and collimated before being directed at the droplet field generating a hologram on the camera sensor. Holograms are recorded at 1400 frames per second (fps) with a 1 $\mu$s exposure time, producing images with a spatial resolution of 40.70 $\mu$m/pixel. Figure S1 in the supplementary information outlines the various steps involved in digital in-line holography processing. The processing method consists of three major steps: (i) holographic reconstruction; (ii) network training; and (iii) post-processing of the holograms. The first step involves subtracting the background image from the recorded holograms, followed by an intensity normalization process to remove noise and correct uneven illumination. Once the holograms are pre-processed, numerical reconstruction is performed using the Rayleigh–Sommerfeld equation to obtain the 3D optical field. The in-plane intensity information is obtained from the magnitude of the complex optical field in the corresponding plane. The reconstruction is performed in a series of planes with spacing $100 ~ \mu\text{m}$ across the test volume centered at the nozzle axis. The dimension of the reconstructed volume is $45~ \text{mm} \times 40 ~\text{mm} \times 31.3  ~ \text{mm}$. 

The primary challenge in the digital inline holography (DIH) technique is the accurate segmentation of particles after hologram reconstruction. To address this, we use U-Net neural network architecture for image segmentation \citep{ade2024application}. The network is initially trained on 500 synthetic holograms and 70 experimental holograms, along with their corresponding binary masks (ground truth). For experimental images, ground truth binary masks are manually annotated using local thresholding. This dataset, which spans various cases, enables the use of a single set of training weights across all experiments. To enhance model robustness and minimize the need for a large dataset, data augmentation techniques, including rigid and elastic deformations of the ground truth images, are applied prior to training \citep{Falk2019}. A comprehensive illustration of the experimental methodology, including shadowgraphy and digital in-line holography, is available in \cite{ade2024prf,ade2023size,ade2022droplet}. Figure S2 (Supplementary information) illustrates the numerical reconstruction of a typical recorded hologram at various depths. The machine-learning approach is employed to estimate the drop size distribution from the images obtained using the in-line holography technique \citep{ade2024application}.  Five repetitions for each Weber number have been considered in the present study to ensure reliable and consistent results.

\section{Fragmentation of raindrop at different Weber numbers}

\begin{figure*}
\centering
\includegraphics[width=0.8\textwidth]{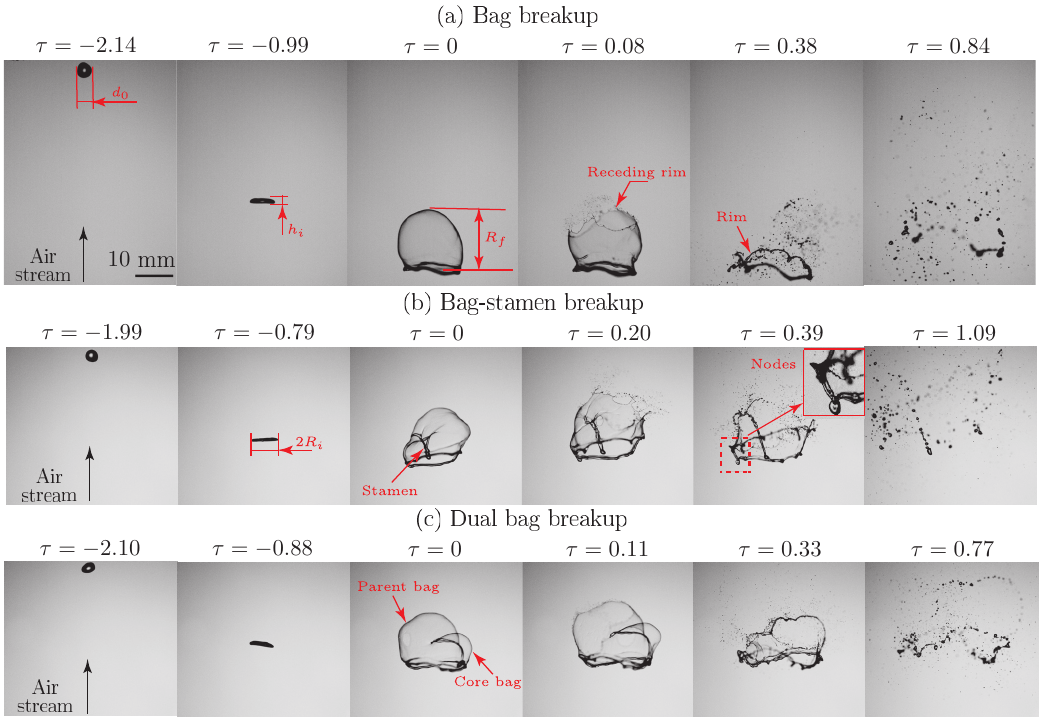}
\caption{Temporal evolution of the droplet breakup dynamics for (a) $\We = 9.38$ (bag breakup), (b) $\We = 16.9$ (bag-stamen breakup) and (c) $\We = 18.9$ (dual bag breakup). The dimensionless time ($\tau = Ut\sqrt{\rho_a/\rho_w}/d_{0}$), wherein $U$ is the average velocity of the airstream, $t$ is the physical time, $\rho_a$ is the density of air, $\rho_w$ is the density of water and $d_{0}$ is the initial droplet diameter. The instant $\tau = 0$ represents the onset of the breakup of the droplet. In the experiments, $d_0=3.6 \pm 0.06$ mm and $\We = 9.38$, 16.9 and 18.9 correspond to airstream velocity $U=12.44$ m/s, $U=16.7$ m/s and $U=17.68$ m/s, respectively.}
\label{temporal_Bag_breakup}
\end{figure*}

Figure \ref{temporal_Bag_breakup} illustrates various droplet breakup phenomena in a counter-current airstream at different Weber numbers, as observed through shadowgraphy. The figure depicts the temporal evolution of three breakup modes: single-bag, bag-stamen, and dual-bag, corresponding to $\We = 9.38$, $\We = 16.9$, and $\We = 18.9$, respectively. The results are presented in terms of the dimensionless time, $\tau = U t \sqrt{\rho_a / \rho_w} / d_0$, which represents the ratio of physical time to inertial time. It is introduced to characterize the rapid timescale associated with droplet deformation and fragmentation processes in an airstream \citep{nicholls1969aerodynamic}. Here, $t$ is the dimensional time in seconds, $\rho_w = 998$ kg/m$^3$ is the density of water and $\tau = 0$ represents the onset of breakup. In a counter-ascending airstream, the single-bag breakup (figure \ref{temporal_Bag_breakup}a) starts with the droplet deforming into an oblate spheroid, resembling a disk shape. As the time progresses, this deformation leads to the formation of a hollow bag attached to a toroidal ring, driven by the pressure difference between the front and back sides of the deformed disk. Eventually, the bag ruptures due to Rayleigh-Taylor (RT) instability, while the subsequent breakup of the toroidal ring is driven by Rayleigh-Plesset (RP) instability. As the Weber number increases, the droplet exhibits a bag-stamen morphology (figure \ref{temporal_Bag_breakup}b). This breakup phenomenon represents a transitional regime in which a bag forms around a slower-moving central core, known as the stamen. This mode arises when both aerodynamic effects and RT instability are significant \citep{guildenbecher2009secondary}. Initially, the bag forms similarly to the normal bag breakup; however, the downstream motion of the central portion leads to the stamen structure. The interplay between RT instabilities and aerodynamic drag generates surface disturbances that are amplified by the airflow, ultimately causing the stamen to break apart. At even higher Weber numbers, the droplet undergoes a dual-bag fragmentation process, involving more intricate dynamics (figure \ref{temporal_Bag_breakup}c). The droplet deforms into a convex shape, leading to the simultaneous inflation of the first bag while a second bag begins to form around the core droplet \citep{ade2023size}. The continuous upward flow sustains the inflation of the first bag and promotes the formation of the second bag by flattening and elongating the core droplet. This persistent aerodynamic action causes the distinct breakup of both bags, making the dual-bag breakup significantly more complex than the single-bag and bag-stamen fragmentation processes. 

\begin{figure*}
\centering
\includegraphics[width=0.7\textwidth]{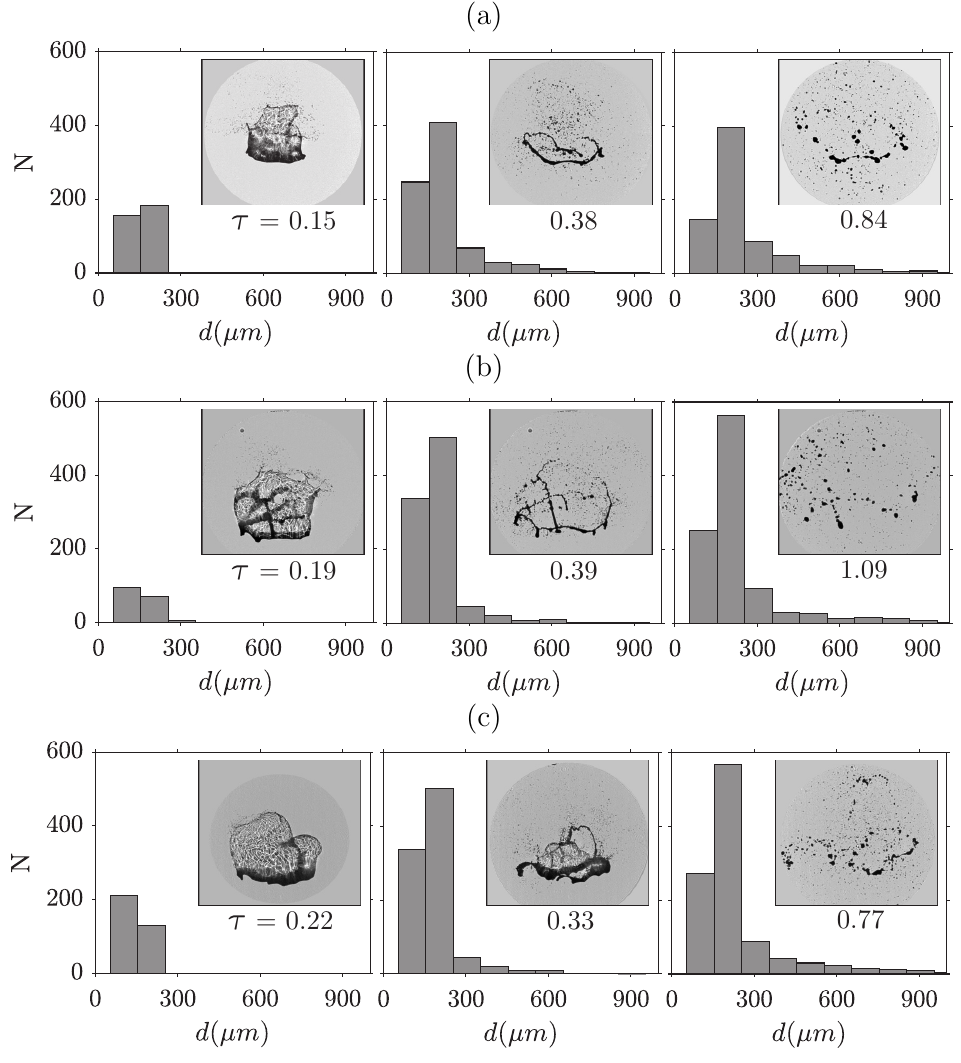}
\caption{Temporal evolution of the droplet size distribution, illustrating droplet counts $N$ as a function of droplet diameter $d$ for (a) $\We = 9.38$ (bag breakup process), (b) $\We = 16.9$ (bag-stamen breakup process), and (c) $\We = 18.9$ (dual-bag breakup process). The inset in each panel displays the recorded holograms obtained using in-line holography.}
\label{number_counts}
\end{figure*}

Figure \ref{number_counts}(a-c) depict the temporal evolution of the recorded holograms obtained from in-line holography, along with the resulting droplet counts for $\We = 9.38$, 16.9 and 18.9, respectively. Figure \ref{number_counts}(a) illustrates that in the early stages, the bag ruptures, generating tiny droplets ($50~\mu$m –$250~\mu$m). As the ruptured sheet retracts, it detaches nodes, increasing the count of smaller droplets. By $\tau = 0.84$, the rim and nodes disintegrate, producing larger droplets ($>300~\mu$m). In the bag-stamen breakup (Figure \ref{number_counts}(b)), fewer tiny droplets ($d < 300~\mu$m) form initially, differing from normal bag breakup. Subsequent fragmentation at $\tau = 0.39$ results in droplets ranging from $50 ~\mu$m–$600~\mu$m. At later stages, the stamen, rim, and nodes disintegrate, creating a broader distribution ($50 ~\mu$m–$1000~\mu$m). For the dual-bag case, simultaneous bag rupture initially generates more child droplets, but the later size distributions align with those of the bag-stamen breakup. A theoretical analysis of these results to predict the DSD will be presented in the following section.

\section{Theoretical modelling}

We employ a two-parameter gamma distribution relationship to present the number size distribution, $P(x)$, of child droplets, which is expressed as \citep{jackiw2022prediction}:
\begin{equation}
P(x = d/d_{avg}) = \frac{x^{\alpha - 1} e^{-x / \beta}}{\beta^\alpha \Gamma(\alpha)}.  \label{eq:gamma_prob_density}
\end{equation}
Here, $\Gamma (\alpha)$ denotes the gamma function, and $\alpha = \left(\bar{x}/s\right)^2$ and $\beta = s^2/\bar{x}$ are the shape and rate parameters, respectively, where $\bar{x}$ and $s$ represent the mean and standard deviation of the distribution, respectively. Additionally, $d_{avg}$ in the distribution is determined from the arithmetic mean of all the characteristic sizes. In the single-parameter gamma size distribution used by \citet{Villermaux2009single}, $\alpha = \beta$. In our study, these parameters are calculated from the mean and standard deviation of the eleven characteristic sizes observed from various instabilities in the bag, rim, and node breakup modes. The eleven characteristic sizes, consisting of four sizes associated with bag breakup, four with rim breakup, and three with node breakup, are discussed in detail below.

{\it Bag rupture:} Droplet breakup starts with the rupture of the bag, where its thickness ($h$) is the first characteristic size, given by \citet{culick1960comments}:
\begin{equation} 
h = \frac{2\sigma}{\rho_w u_{rr}^2}, 
\end{equation}
with $u_{rr}$ being the experimentally determined retraction speed. For water, $h \approx 2.3 \pm 1.2$ $\mu$m when $\We < 30$ \citep{jackiw2022prediction}. After rupture, the edge of the bag forms a receding rim, with thickness ($b_{rr}$) as the second characteristic size, given by:
\begin{equation} 
b_{rr} = \sqrt{\frac{\sigma}{\rho_w a_c}}, 
\end{equation}
where $a_c$ is the acceleration of the rim \citep{wang2018universal}. The rim then undergoes the RP instability, leading to the third characteristic size, $d_{RP,B}$ \citep{jackiw2022prediction}:
\begin{equation} 
d_{RP,B} = 1.89 b_{rr}. 
\end{equation}
The formation and breakup of ligaments from the edge of the bag, driven by nonlinear instability, result in the creation of tiny satellite droplets. This fourth characteristic size ($d_{sat,B}$) can be calculated as \citep{keshavarz2020rotary}:
\begin{equation} 
d_{sat,B} = \frac{d_{RP,B}}{\sqrt{2 + {3 Oh_{rr}/\sqrt{2}}}}, 
\end{equation}
where $Oh_{rr}=\mu_w/\sqrt{\rho_w \sigma b_{rr}}$ is the Ohnesorge number based on the receding rim thickness, wherein $\mu_w$ is the viscosity of water.

{\it Rim fragmentation:} The rim fragmentation begins after the bag bursts, primarily driven by the RP instability. The first characteristic breakup size due to this instability is given by
\begin{equation} 
d_R = 1.89 h_f. 
\end{equation}
Here, $h_{f} = h_{i}\sqrt{R_{i}/R_{f}}$ represents the final rim thickness, wherein $R_i$ denotes the maximum disk radius, and $R_{f}$ is the radius of the bag at the instant when it bursts (figure \ref{temporal_Bag_breakup}). The second characteristic size due to receding rim collision ($d_{rr}$) is given by \citep{jackiw2022prediction}:
\begin{equation} 
d_{rr} = d_0 \left[ \frac{3}{2} \left( \frac{h_f}{d_0} \right)^2 \frac{\lambda_{rr}}{d_0} \right]^{1/3}, 
\end{equation}
where $\lambda_{rr} = 4.5b_{rr}$ is the wavelength of the receding rim instability.

\begin{figure}
\centering
\includegraphics[width=0.45\textwidth]{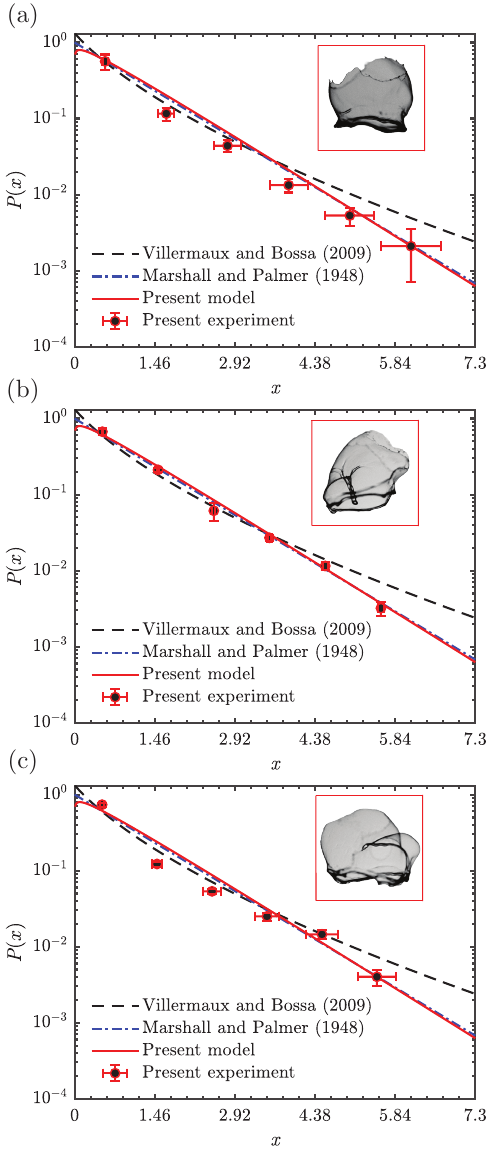}
\caption{Drop size distribution for different breakup phenomena. Variations of $P(x)$ with $x = d/d_{avg}$ obtained from our experiments are shown for (a) $\We = 9.38$ (bag breakup), (b) $\We = 16.9$ (bag-stamen breakup), and (c) $\We = 18.9$ (dual bag breakup). The present theoretical model, along with the models by \citet{Villermaux2009single} and \citet{marshall1948distribution}, are also depicted. The present theoretical model closely predicts the experimental results and aligns well with the Marshall-Palmer relationship. The error bars indicate the standard deviation associated with the five independent repetitions for each Weber number.}
\label{comparison_with_villermaux}
\end{figure}

The final mechanism involves satellite droplets formed from liquid ligaments near the pinch-off point, affecting both receding rim collision and the Rayleigh-Plateau breakup. The characteristic sizes of these satellite droplets are \citep{keshavarz2020rotary}:
\begin{eqnarray} d_{sat,R} = \frac{d_R}{\sqrt{2 + 3 Oh_R / \sqrt{2}}}, \\ 
d_{sat,rr} = \frac{d_{rr}}{\sqrt{2 + 3 Oh_R / \sqrt{2}}}, 
\end{eqnarray}
where $Oh_{R} = {\mu_w/\sqrt{\rho_w h_{f}^3 \sigma}}$ is the Ohnesorge number based on the final rim thickness. More details can be found in  \citet{jackiw2021aerodynamic,jackiw2022prediction,ade2022droplet}.

{\it Node breakup:} Node (finger-like structures) breakup occurs due to the RP and RT instabilities, as detailed by \citet{zhao2010morphological} and \citet{kirar2022experimental}. The size $d_{N}$ of child droplets from node breakup is given by \citep{jackiw2022prediction}:
\begin{equation} \label{j:eq9}
\frac{d_{N}}{d_{0}}=\left [ \frac{3}{2}\left ( \frac{h_i}{d_{0}} \right )^{2}\frac{\lambda_{RT} }{d_{0}} V_n \right ]^{1/3},
\end{equation}
where $h_i$ is the disk thickness, $\lambda_{RT}$ is the RT instability wavelength, and $V_n$ is the node volume fraction relative to the disk. \citet{jackiw2022prediction} estimated the values of $V_n$ as 0.2, 0.4, and 1 for minimum, mean, and maximum sizes, respectively. The node droplets exhibit three characteristic sizes, which can be determined using these three values of $V_n$. Thus, there are eleven characteristic breakup sizes: four from bag breakup, four from rim breakup, and three from node breakup. All characteristic sizes are measured in $\mu$m.

Figure \ref{comparison_with_villermaux} illustrates the variation of the number size distribution of child droplets, $P(x)$, with $x = d/d_{avg}$ from our experiments for different Weber numbers, each corresponding to distinct breakup phenomena. The results show that our model, which incorporates all characteristic sizes resulting from various breakup modes, accurately predicts the experimental observations. We have conducted a large number of experiments considering different initial droplet diameters, but for clarity, we have presented the data only for $d_0 = 3.6 \pm 0.08$ mm. All the characteristic sizes mentioned earlier are utilized to estimate the mean $(\bar{x})$ and standard deviation $(s)$, which are then used to derive the parameters $\alpha$ and $\beta$ for specific breakup phenomena. Using the values of $\alpha$ and $\beta$ for different Weber numbers, we estimate the number size distribution, $P(x)$, from Eq. (\ref{eq:gamma_prob_density}). Moreover, $d_{avg}$, an unknown parameter in \citet{Villermaux2009single}, is now directly calculated from the characteristic sizes obtained from our theoretical model. Figure \ref{comparison_with_villermaux} also depicts a comparison of our results with the theoretical model by \citet{Villermaux2009single} and the empirical relationship by Marshall and Palmer, which is based on direct measurements of drop size distribution. It is evident that while our experimental results and theoretical model align closely with the findings of \citet{marshall1948distribution}, the model by \citet{Villermaux2009single} shows significant deviations, particularly for larger droplets ($d/d_{avg} \gtrapprox 3$) across all breakup scenarios considered. Specifically, the deviations of the model proposed by \citet{Villermaux2009single} with respect to the distribution of \citet{marshall1948distribution} for bag, bag-stamen, and dual-bag breakup cases are 25\%, 24.2\%, and 27.1\%, respectively, at $d/d_{avg}=7.3$. The corresponding deviations of our model from that of \citet{marshall1948distribution} are 7.6\%, 7.1\%, and 6.6\%, respectively, which quantitatively justifies the robustness of the present model. Furthermore, we compute the integral $\int x^{\frac{7}{2}} P(x) \d x$ in Eq. (\ref{eq:davg_rainfall_rate}), leading to the relationship between $d_{avg}$ and rainfall rate $\R$ as
\begin{equation} 
d_{avg}^{-1} = 40 \R^{-0.21}, \label{mod_eq:rainfall_rate} 
\end{equation} 
such that $\Lambda=40$ and $a=0.21$. This relationship closely matches the one proposed by Marshall and Palmer for steady rain \citep{marshall1948distribution}, suggesting that all instability mechanisms involved in the fragmentation process contribute to the observed drop size distribution. Next, we compare our theoretically and experimentally obtained number size distributions with the in situ measurements.

\section{Aerial in situ measurements} 

\begin{figure*}
\centering
\includegraphics[width=0.75\textwidth]{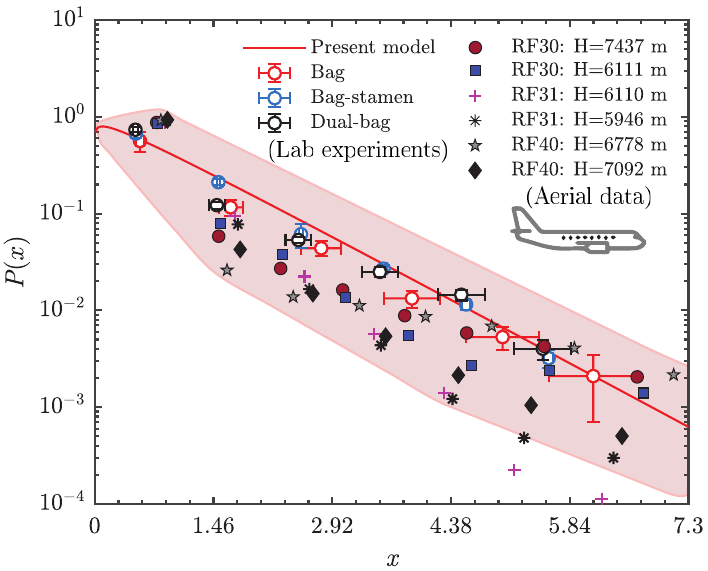}
\caption{{\bf Aerial in situ measurements using research flights at different altitudes $(H)$.} A comparison of the variation of $P(x)$ as a function of $x = d/d_{avg}$ is presented, incorporating data from the research flights (RF), along with our laboratory experimental results and theoretical predictions. The shaded envelope represents the complete range of scattered data points from aerial measurements at various altitudes and experimental results across different Weber numbers, relative to our theoretical model.}
\label{flight_data}
\end{figure*}

The Aero Commander 690A was utilized as an airborne research platform for the Cloud Aerosol Interaction and Precipitation Enhancement Experiment (CAIPEEX Phase II), a research mission aimed at documenting aerosol-cloud interactions. The precipitation data for this study were collected in both continental and maritime clouds. Further details of the CAIPEEX measurements can be found in \citet{prabhakaran2023caipeex}.

The research flight was equipped with several advanced sensors and was capable of withstanding severe icing, turbulence, high temperatures, and high altitudes \citep{patade2015particle}. It featured six Particle Measuring Systems (PMS), including the Forward Scatter Spectrometer Probe (FSSP-100; \citet{dye1984evaluation}), the DMT (Droplet Measurement Technologies Inc.) Cloud Aerosol Spectrometer (CAS; \citet{baumgardner2001cloud}), the DMT Cloud Imaging Probe (CIP), the DMT Precipitation Imaging Probe (PIP), the DMT Liquid Water Content (LWC) hot-wire probe, and the King LWC hot-wire probe. Although the flights used in CAIPEEX were equipped with various sensors to measure different parameters, the CIP and PIP data are used to analyze the raindrop size distribution in the present study. \ks{The CIP detects droplets ranging from 15 $\mu$m to 913 $\mu$m with a resolution of 15 $\mu$m, while the PIP measures sizes from 100 $\mu$m to 6000 $\mu$m with a resolution of 10 $\mu$m \citep{baumgardner2001cloud}.} Both instruments featured anti-shattering tips to reduce artifacts and ensure accuracy. Measurements were recorded at 1 Hz, and data were processed using the System for Optical Array Probe Data Analysis (SODA) code developed at the National Center for Atmospheric Research (NCAR). Further details on the instruments, operational ranges, and data processing are available in \citet{patade2016observational,patade2015particle}.

The precipitation data at various altitudes were obtained from three research flights, as listed below.
\begin{itemize} 
\item RF30: Launched on October 19, 2011, from Bhubaneswar, India, this flight operated from 7:47 AM to 11:29 AM, reaching an altitude of 7468 meters, with temperatures ranging from -12.56 $^\circ$C to 38.53 $^\circ$C and static pressures between 41.45 kPa and 100.85 kPa. 
\item RF31: On October 20, 2011, this flight travelled from Bhubaneswar to Vijayawada, operating from 6:13 AM to 9:31 AM, reaching 6145 meters with temperatures from -6.12 $^\circ$C to 40.29 $^\circ$C and static pressures between 48.87 kPa and 101.04 kPa. 
\item RF40: Conducted on October 27, 2011, from Vijayawada to Hyderabad, this flight reached an altitude of 7448 meters, with temperatures between -11.57 $^\circ$C and 34.25 $^\circ$C and static pressures between 41.55 kPa and 95.83 kPa. 
\end{itemize}
\ks{Figure 4 of \citet{patade2015particle} presents the variations in droplet number concentration per unit size per unit volume, $C(d)$, as a function of droplet size, $d$, based on measurements from three research flights (RF30, RF31, and RF40) conducted at different altitudes, $H$. These data were utilized for comparison with the current laboratory experimental results and theoretical predictions.} In our study, Figure \ref{flight_data} illustrates the variation of $P(x)$ as a function of $x = d/d_{avg}$ obtained from these precipitation data. The results from laboratory experiments and theoretical predictions (Eq. \ref{eq:gamma_prob_density}) are also depicted in Figure \ref{flight_data}, showing a satisfactory alignment with the in situ aerial raindrop size measurements. This agreement suggests that laboratory experiments and theoretical models can effectively predict rainfall under various conditions.

\section{Concluding remarks} 

In the present study, laboratory experiments utilizing Shadowgraphy, digital in-line holography, and theoretical modelling are performed to determine the raindrop size distribution (DSD) resulting from distinct fragmentation processes under various upward airstreams, characterized by the Weber number. For the theoretical modelling, we employ a two-parameter gamma distribution to estimate the average droplet diameter based on characteristic sizes, which are often treated as assumed input parameters for different rainfall conditions. Additionally, in situ DSD measurements at various altitudes were obtained through research flights equipped with advanced sensors, further validating our rainfall model. \cite{Villermaux2009single} also emphasized the importance of measuring in situ raindrop size distributions during precipitation and comparing them with laboratory experiments. The slight scattering of the in situ data at larger droplet sizes in Figure \ref{flight_data} could be due to the presence of snowflakes mixed with raindrops in the flight data. 
Moreover, actual atmospheric conditions, such as local temperature and humidity, aerosol content at different altitudes, and other microphysical processes \citep{raut2021microphysical} like collision-coalescence, collisional breakup \citep{srivastava1978parameterization, low1982collision, mcfarquhar2004new, straub2010numerical, testik2017first}, evaporation, and condensation may influence the DSD, warranting further investigation in future studies.

\acknowledgments
K.C.S. and L.D.C. thank IIT Hyderabad for financial support through Grant No. IITH/CHE/F011/SOCH1. S.S.A. thanks the PMRF Fellowship. The CAIPEEX project was funded by the Ministry of Earth Sciences (Government of India) and conducted by the Indian Institute of Tropical Meteorology (IITM), Pune. We also acknowledge the support of Neelam Malap for her assistance in explaining the CAIPEEX data.


\begin{thebibliography}{41}%
\makeatletter
\providecommand \@ifxundefined [1]{%
 \@ifx{#1\undefined}
}%
\providecommand \@ifnum [1]{%
 \ifnum #1\expandafter \@firstoftwo
 \else \expandafter \@secondoftwo
 \fi
}%
\providecommand \@ifx [1]{%
 \ifx #1\expandafter \@firstoftwo
 \else \expandafter \@secondoftwo
 \fi
}%
\providecommand \natexlab [1]{#1}%
\providecommand \enquote  [1]{``#1''}%
\providecommand \bibnamefont  [1]{#1}%
\providecommand \bibfnamefont [1]{#1}%
\providecommand \citenamefont [1]{#1}%
\providecommand \href@noop [0]{\@secondoftwo}%
\providecommand \href [0]{\begingroup \@sanitize@url \@href}%
\providecommand \@href[1]{\@@startlink{#1}\@@href}%
\providecommand \@@href[1]{\endgroup#1\@@endlink}%
\providecommand \@sanitize@url [0]{\catcode `\\12\catcode `\$12\catcode
  `\&12\catcode `\#12\catcode `\^12\catcode `\_12\catcode `\%12\relax}%
\providecommand \@@startlink[1]{}%
\providecommand \@@endlink[0]{}%
\providecommand \url  [0]{\begingroup\@sanitize@url \@url }%
\providecommand \@url [1]{\endgroup\@href {#1}{\urlprefix }}%
\providecommand \urlprefix  [0]{URL }%
\providecommand \Eprint [0]{\href }%
\providecommand \doibase [0]{http://dx.doi.org/}%
\providecommand \selectlanguage [0]{\@gobble}%
\providecommand \bibinfo  [0]{\@secondoftwo}%
\providecommand \bibfield  [0]{\@secondoftwo}%
\providecommand \translation [1]{[#1]}%
\providecommand \BibitemOpen [0]{}%
\providecommand \bibitemStop [0]{}%
\providecommand \bibitemNoStop [0]{.\EOS\space}%
\providecommand \EOS [0]{\spacefactor3000\relax}%
\providecommand \BibitemShut  [1]{\csname bibitem#1\endcsname}%
\let\auto@bib@innerbib\@empty
\bibitem [{\citenamefont {Marinescu}\ \emph {et~al.}(2020)\citenamefont
  {Marinescu}, \citenamefont {Kennedy}, \citenamefont {Bell}, \citenamefont
  {Drager}, \citenamefont {Grant}, \citenamefont {Freeman},\ and\ \citenamefont
  {van~den Heever}}]{marinescu2020updraft}%
  \BibitemOpen
  \bibfield  {author} {\bibinfo {author} {\bibfnamefont {P.~J.}\ \bibnamefont
  {Marinescu}}, \bibinfo {author} {\bibfnamefont {P.~C.}\ \bibnamefont
  {Kennedy}}, \bibinfo {author} {\bibfnamefont {M.~M.}\ \bibnamefont {Bell}},
  \bibinfo {author} {\bibfnamefont {A.~J.}\ \bibnamefont {Drager}}, \bibinfo
  {author} {\bibfnamefont {L.~D.}\ \bibnamefont {Grant}}, \bibinfo {author}
  {\bibfnamefont {S.~W.}\ \bibnamefont {Freeman}}, \ and\ \bibinfo {author}
  {\bibfnamefont {S.~C.}\ \bibnamefont {van~den Heever}},\ }\bibfield  {title}
  {\enquote {\bibinfo {title} {Updraft vertical velocity observations and
  uncertainties in high plains supercells using radiosondes and radars},}\
  }\href@noop {} {\bibfield  {journal} {\bibinfo  {journal} {Mon. Weather
  Rev.}\ }\textbf {\bibinfo {volume} {148}},\ \bibinfo {pages} {4435--4452}
  (\bibinfo {year} {2020})}\BibitemShut {NoStop}%
\bibitem [{\citenamefont {Musil}, \citenamefont {Heymsfield},\ and\
  \citenamefont {Smith}(1986)}]{musil1986microphysical}%
  \BibitemOpen
  \bibfield  {author} {\bibinfo {author} {\bibfnamefont {D.~J.}\ \bibnamefont
  {Musil}}, \bibinfo {author} {\bibfnamefont {A.~J.}\ \bibnamefont
  {Heymsfield}}, \ and\ \bibinfo {author} {\bibfnamefont {P.~L.}\ \bibnamefont
  {Smith}},\ }\bibfield  {title} {\enquote {\bibinfo {title} {Microphysical
  characteristics of a well-developed weak echo region in a high plains
  supercell thunderstorm},}\ }\href@noop {} {\bibfield  {journal} {\bibinfo
  {journal} {J. Climate Appl. Meteor.}\ }\textbf {\bibinfo {volume} {25}},\
  \bibinfo {pages} {1037--1051} (\bibinfo {year} {1986})}\BibitemShut {NoStop}%
\bibitem [{\citenamefont {Uijlenhoet}(2001)}]{uijlenhoet2001raindrop}%
  \BibitemOpen
  \bibfield  {author} {\bibinfo {author} {\bibfnamefont {R.}~\bibnamefont
  {Uijlenhoet}},\ }\bibfield  {title} {\enquote {\bibinfo {title} {Raindrop
  size distributions and radar reflectivity--rain rate relationships for radar
  hydrology},}\ }\href@noop {} {\bibfield  {journal} {\bibinfo  {journal}
  {Hydrol. Earth Syst. Sci.}\ }\textbf {\bibinfo {volume} {5}},\ \bibinfo
  {pages} {615--628} (\bibinfo {year} {2001})}\BibitemShut {NoStop}%
\bibitem [{\citenamefont {Laws}\ and\ \citenamefont
  {Parsons}(1943)}]{laws1943relation}%
  \BibitemOpen
  \bibfield  {author} {\bibinfo {author} {\bibfnamefont {J.~O.}\ \bibnamefont
  {Laws}}\ and\ \bibinfo {author} {\bibfnamefont {D.~A.}\ \bibnamefont
  {Parsons}},\ }\bibfield  {title} {\enquote {\bibinfo {title} {The relation of
  raindrop-size to intensity},}\ }\href@noop {} {\bibfield  {journal} {\bibinfo
   {journal} {Trans. Am. Geophys. Union.}\ }\textbf {\bibinfo {volume} {24}},\
  \bibinfo {pages} {452--460} (\bibinfo {year} {1943})}\BibitemShut {NoStop}%
\bibitem [{\citenamefont {Marshall}\ and\ \citenamefont
  {Palmer}(1948)}]{marshall1948distribution}%
  \BibitemOpen
  \bibfield  {author} {\bibinfo {author} {\bibfnamefont {J.~S.}\ \bibnamefont
  {Marshall}}\ and\ \bibinfo {author} {\bibfnamefont {W.~M.}\ \bibnamefont
  {Palmer}},\ }\bibfield  {title} {\enquote {\bibinfo {title} {The distribution
  of raindrops with size},}\ }\href@noop {} {\bibfield  {journal} {\bibinfo
  {journal} {J. Meteorol.}\ }\textbf {\bibinfo {volume} {5}},\ \bibinfo {pages}
  {165--166} (\bibinfo {year} {1948})}\BibitemShut {NoStop}%
\bibitem [{\citenamefont {Yau}\ and\ \citenamefont
  {Rogers}(1996)}]{yau1996short}%
  \BibitemOpen
  \bibfield  {author} {\bibinfo {author} {\bibfnamefont {M.~K.}\ \bibnamefont
  {Yau}}\ and\ \bibinfo {author} {\bibfnamefont {R.~R.}\ \bibnamefont
  {Rogers}},\ }\href@noop {} {\emph {\bibinfo {title} {A short course in cloud
  physics}}}\ (\bibinfo  {publisher} {Elsevier},\ \bibinfo {year}
  {1996})\BibitemShut {NoStop}%
\bibitem [{\citenamefont {Villermaux}\ and\ \citenamefont
  {Bossa}(2009)}]{Villermaux2009single}%
  \BibitemOpen
  \bibfield  {author} {\bibinfo {author} {\bibfnamefont {E.}~\bibnamefont
  {Villermaux}}\ and\ \bibinfo {author} {\bibfnamefont {B.}~\bibnamefont
  {Bossa}},\ }\bibfield  {title} {\enquote {\bibinfo {title} {Single-drop
  fragmentation determines size distribution of raindrops},}\ }\href@noop {}
  {\bibfield  {journal} {\bibinfo  {journal} {Nat. Phys.}\ }\textbf {\bibinfo
  {volume} {5}},\ \bibinfo {pages} {697--702} (\bibinfo {year}
  {2009})}\BibitemShut {NoStop}%
\bibitem [{\citenamefont {Mezhericher}\ and\ \citenamefont
  {Stone}(2022)}]{mezhericher2022size}%
  \BibitemOpen
  \bibfield  {author} {\bibinfo {author} {\bibfnamefont {M.}~\bibnamefont
  {Mezhericher}}\ and\ \bibinfo {author} {\bibfnamefont {H.~A.}\ \bibnamefont
  {Stone}},\ }\bibfield  {title} {\enquote {\bibinfo {title} {Size distribution
  of raindrops},}\ }\href@noop {} {\bibfield  {journal} {\bibinfo  {journal}
  {arXiv preprint arXiv:2204.03151}\ } (\bibinfo {year} {2022})}\BibitemShut
  {NoStop}%
\bibitem [{\citenamefont {Pilch}\ and\ \citenamefont
  {Erdman}(1987)}]{pilch1987use}%
  \BibitemOpen
  \bibfield  {author} {\bibinfo {author} {\bibfnamefont {M.}~\bibnamefont
  {Pilch}}\ and\ \bibinfo {author} {\bibfnamefont {C.~A.}\ \bibnamefont
  {Erdman}},\ }\bibfield  {title} {\enquote {\bibinfo {title} {Use of breakup
  time data and velocity history data to predict the maximum size of stable
  fragments for acceleration-induced breakup of a liquid drop},}\ }\href@noop
  {} {\bibfield  {journal} {\bibinfo  {journal} {Int. J. Multiphase Flow}\
  }\textbf {\bibinfo {volume} {13}},\ \bibinfo {pages} {741--757} (\bibinfo
  {year} {1987})}\BibitemShut {NoStop}%
\bibitem [{\citenamefont {Cao}\ \emph {et~al.}(2007)\citenamefont {Cao},
  \citenamefont {Sun}, \citenamefont {Li}, \citenamefont {Liu},\ and\
  \citenamefont {Yu}}]{cao2007new}%
  \BibitemOpen
  \bibfield  {author} {\bibinfo {author} {\bibfnamefont {X.~K.}\ \bibnamefont
  {Cao}}, \bibinfo {author} {\bibfnamefont {Z.~G.}\ \bibnamefont {Sun}},
  \bibinfo {author} {\bibfnamefont {W.~F.}\ \bibnamefont {Li}}, \bibinfo
  {author} {\bibfnamefont {H.~F.}\ \bibnamefont {Liu}}, \ and\ \bibinfo
  {author} {\bibfnamefont {Z.~H.}\ \bibnamefont {Yu}},\ }\bibfield  {title}
  {\enquote {\bibinfo {title} {A new breakup regime of liquid drops identified
  in a continuous and uniform air jet flow},}\ }\href@noop {} {\bibfield
  {journal} {\bibinfo  {journal} {Phys. Fluids}\ }\textbf {\bibinfo {volume}
  {19}},\ \bibinfo {pages} {057103} (\bibinfo {year} {2007})}\BibitemShut
  {NoStop}%
\bibitem [{\citenamefont {Guildenbecher}, \citenamefont {L{\'o}pez-Rivera},\
  and\ \citenamefont {Sojka}(2009)}]{guildenbecher2009secondary}%
  \BibitemOpen
  \bibfield  {author} {\bibinfo {author} {\bibfnamefont {D.~R.}\ \bibnamefont
  {Guildenbecher}}, \bibinfo {author} {\bibfnamefont {C.}~\bibnamefont
  {L{\'o}pez-Rivera}}, \ and\ \bibinfo {author} {\bibfnamefont {P.~E.}\
  \bibnamefont {Sojka}},\ }\bibfield  {title} {\enquote {\bibinfo {title}
  {Secondary atomization},}\ }\href@noop {} {\bibfield  {journal} {\bibinfo
  {journal} {Exp. Fluids}\ }\textbf {\bibinfo {volume} {46}},\ \bibinfo {pages}
  {371--402} (\bibinfo {year} {2009})}\BibitemShut {NoStop}%
\bibitem [{\citenamefont {Suryaprakash}\ and\ \citenamefont
  {Tomar}(2019)}]{suryaprakash2019secondary}%
  \BibitemOpen
  \bibfield  {author} {\bibinfo {author} {\bibfnamefont {R.}~\bibnamefont
  {Suryaprakash}}\ and\ \bibinfo {author} {\bibfnamefont {G.}~\bibnamefont
  {Tomar}},\ }\bibfield  {title} {\enquote {\bibinfo {title} {Secondary breakup
  of drops},}\ }\href@noop {} {\bibfield  {journal} {\bibinfo  {journal} {J.
  Indian Inst. Sci.}\ }\textbf {\bibinfo {volume} {99}},\ \bibinfo {pages}
  {77--91} (\bibinfo {year} {2019})}\BibitemShut {NoStop}%
\bibitem [{\citenamefont {Soni}\ \emph {et~al.}(2020)\citenamefont {Soni},
  \citenamefont {Kirar}, \citenamefont {Kolhe},\ and\ \citenamefont
  {Sahu}}]{soni2020deformation}%
  \BibitemOpen
  \bibfield  {author} {\bibinfo {author} {\bibfnamefont {S.~K.}\ \bibnamefont
  {Soni}}, \bibinfo {author} {\bibfnamefont {P.~K.}\ \bibnamefont {Kirar}},
  \bibinfo {author} {\bibfnamefont {P.}~\bibnamefont {Kolhe}}, \ and\ \bibinfo
  {author} {\bibfnamefont {K.~C.}\ \bibnamefont {Sahu}},\ }\bibfield  {title}
  {\enquote {\bibinfo {title} {Deformation and breakup of droplets in an
  oblique continuous air stream},}\ }\href@noop {} {\bibfield  {journal}
  {\bibinfo  {journal} {Int. J. Multiphase Flow}\ }\textbf {\bibinfo {volume}
  {122}},\ \bibinfo {pages} {103141} (\bibinfo {year} {2020})}\BibitemShut
  {NoStop}%
\bibitem [{\citenamefont {Kirar}\ \emph {et~al.}(2022)\citenamefont {Kirar},
  \citenamefont {Soni}, \citenamefont {Kolhe},\ and\ \citenamefont
  {Sahu}}]{kirar2022experimental}%
  \BibitemOpen
  \bibfield  {author} {\bibinfo {author} {\bibfnamefont {P.~K.}\ \bibnamefont
  {Kirar}}, \bibinfo {author} {\bibfnamefont {S.~K.}\ \bibnamefont {Soni}},
  \bibinfo {author} {\bibfnamefont {P.~S.}\ \bibnamefont {Kolhe}}, \ and\
  \bibinfo {author} {\bibfnamefont {K.~C.}\ \bibnamefont {Sahu}},\ }\bibfield
  {title} {\enquote {\bibinfo {title} {An experimental investigation of droplet
  morphology in swirl flow},}\ }\href@noop {} {\bibfield  {journal} {\bibinfo
  {journal} {J. Fluid Mech.}\ }\textbf {\bibinfo {volume} {938}},\ \bibinfo
  {pages} {A6} (\bibinfo {year} {2022})}\BibitemShut {NoStop}%
\bibitem [{\citenamefont {Taylor}(1963)}]{taylor1963shape}%
  \BibitemOpen
  \bibfield  {author} {\bibinfo {author} {\bibfnamefont {G.~I.}\ \bibnamefont
  {Taylor}},\ }\bibfield  {title} {\enquote {\bibinfo {title} {The shape and
  acceleration of a drop in a high speed air stream},}\ }\href@noop {}
  {\bibfield  {journal} {\bibinfo  {journal} {The Scientific Papers of G. I.
  Taylor}\ }\textbf {\bibinfo {volume} {3}},\ \bibinfo {pages} {457--464}
  (\bibinfo {year} {1963})}\BibitemShut {NoStop}%
\bibitem [{\citenamefont {Jackiw}\ and\ \citenamefont
  {Ashgriz}(2021)}]{jackiw2021aerodynamic}%
  \BibitemOpen
  \bibfield  {author} {\bibinfo {author} {\bibfnamefont {I.~M.}\ \bibnamefont
  {Jackiw}}\ and\ \bibinfo {author} {\bibfnamefont {N.}~\bibnamefont
  {Ashgriz}},\ }\bibfield  {title} {\enquote {\bibinfo {title} {On aerodynamic
  droplet breakup},}\ }\href@noop {} {\bibfield  {journal} {\bibinfo  {journal}
  {J. Fluid Mech.}\ }\textbf {\bibinfo {volume} {913}},\ \bibinfo {pages} {A33}
  (\bibinfo {year} {2021})}\BibitemShut {NoStop}%
\bibitem [{\citenamefont {Jackiw}\ and\ \citenamefont
  {Ashgriz}(2022)}]{jackiw2022prediction}%
  \BibitemOpen
  \bibfield  {author} {\bibinfo {author} {\bibfnamefont {I.~M.}\ \bibnamefont
  {Jackiw}}\ and\ \bibinfo {author} {\bibfnamefont {N.}~\bibnamefont
  {Ashgriz}},\ }\bibfield  {title} {\enquote {\bibinfo {title} {Prediction of
  the droplet size distribution in aerodynamic droplet breakup},}\ }\href@noop
  {} {\bibfield  {journal} {\bibinfo  {journal} {J. Fluid Mech.}\ }\textbf
  {\bibinfo {volume} {940}},\ \bibinfo {pages} {A17} (\bibinfo {year}
  {2022})}\BibitemShut {NoStop}%
\bibitem [{\citenamefont {Ade}, \citenamefont {Chandrala},\ and\ \citenamefont
  {Sahu}(2023)}]{ade2023size}%
  \BibitemOpen
  \bibfield  {author} {\bibinfo {author} {\bibfnamefont {S.~S.}\ \bibnamefont
  {Ade}}, \bibinfo {author} {\bibfnamefont {L.~D.}\ \bibnamefont {Chandrala}},
  \ and\ \bibinfo {author} {\bibfnamefont {K.~C.}\ \bibnamefont {Sahu}},\
  }\bibfield  {title} {\enquote {\bibinfo {title} {Size distribution of a drop
  undergoing breakup at moderate weber numbers},}\ }\href@noop {} {\bibfield
  {journal} {\bibinfo  {journal} {J. Fluid Mech.}\ }\textbf {\bibinfo {volume}
  {959}},\ \bibinfo {pages} {A38} (\bibinfo {year} {2023})}\BibitemShut
  {NoStop}%
\bibitem [{\citenamefont {Agrawal}\ \emph {et~al.}(2017)\citenamefont
  {Agrawal}, \citenamefont {Premlata}, \citenamefont {Tripathi}, \citenamefont
  {Karri},\ and\ \citenamefont {Sahu}}]{agrawal2017nonspherical}%
  \BibitemOpen
  \bibfield  {author} {\bibinfo {author} {\bibfnamefont {M.}~\bibnamefont
  {Agrawal}}, \bibinfo {author} {\bibfnamefont {A.~R.}\ \bibnamefont
  {Premlata}}, \bibinfo {author} {\bibfnamefont {M.~K.}\ \bibnamefont
  {Tripathi}}, \bibinfo {author} {\bibfnamefont {B.}~\bibnamefont {Karri}}, \
  and\ \bibinfo {author} {\bibfnamefont {K.~C.}\ \bibnamefont {Sahu}},\
  }\bibfield  {title} {\enquote {\bibinfo {title} {Nonspherical liquid droplet
  falling in air},}\ }\href@noop {} {\bibfield  {journal} {\bibinfo  {journal}
  {Phys. Rev. E.}\ }\textbf {\bibinfo {volume} {95}},\ \bibinfo {pages}
  {033111} (\bibinfo {year} {2017})}\BibitemShut {NoStop}%
\bibitem [{\citenamefont {Balla}, \citenamefont {Tripathi},\ and\ \citenamefont
  {Sahu}(2019)}]{balla2019shape}%
  \BibitemOpen
  \bibfield  {author} {\bibinfo {author} {\bibfnamefont {M.}~\bibnamefont
  {Balla}}, \bibinfo {author} {\bibfnamefont {M.~K.}\ \bibnamefont {Tripathi}},
  \ and\ \bibinfo {author} {\bibfnamefont {K.~C.}\ \bibnamefont {Sahu}},\
  }\bibfield  {title} {\enquote {\bibinfo {title} {Shape oscillations of a
  nonspherical water droplet},}\ }\href@noop {} {\bibfield  {journal} {\bibinfo
   {journal} {Phys. Rev. E.}\ }\textbf {\bibinfo {volume} {99}},\ \bibinfo
  {pages} {023107} (\bibinfo {year} {2019})}\BibitemShut {NoStop}%
\bibitem [{\citenamefont {Khain}\ \emph {et~al.}(2013)\citenamefont {Khain},
  \citenamefont {Prabha}, \citenamefont {Benmoshe}, \citenamefont
  {Pandithurai},\ and\ \citenamefont {Ovchinnikov}}]{khain2013mechanism}%
  \BibitemOpen
  \bibfield  {author} {\bibinfo {author} {\bibfnamefont {A.}~\bibnamefont
  {Khain}}, \bibinfo {author} {\bibfnamefont {T.~V.}\ \bibnamefont {Prabha}},
  \bibinfo {author} {\bibfnamefont {N.}~\bibnamefont {Benmoshe}}, \bibinfo
  {author} {\bibfnamefont {G.}~\bibnamefont {Pandithurai}}, \ and\ \bibinfo
  {author} {\bibfnamefont {M.}~\bibnamefont {Ovchinnikov}},\ }\bibfield
  {title} {\enquote {\bibinfo {title} {The mechanism of first raindrops
  formation in deep convective clouds},}\ }\href@noop {} {\bibfield  {journal}
  {\bibinfo  {journal} {J. Geophys. Res. Atmos.}\ }\textbf {\bibinfo {volume}
  {118}},\ \bibinfo {pages} {9123--9140} (\bibinfo {year} {2013})}\BibitemShut
  {NoStop}%
\bibitem [{\citenamefont {Ade}\ \emph {et~al.}(2024{\natexlab{a}})\citenamefont
  {Ade}, \citenamefont {Gupta}, \citenamefont {Chandrala},\ and\ \citenamefont
  {Sahu}}]{ade2024application}%
  \BibitemOpen
  \bibfield  {author} {\bibinfo {author} {\bibfnamefont {S.~S.}\ \bibnamefont
  {Ade}}, \bibinfo {author} {\bibfnamefont {D.}~\bibnamefont {Gupta}}, \bibinfo
  {author} {\bibfnamefont {L.~D.}\ \bibnamefont {Chandrala}}, \ and\ \bibinfo
  {author} {\bibfnamefont {K.~C.}\ \bibnamefont {Sahu}},\ }\bibfield  {title}
  {\enquote {\bibinfo {title} {Application of deep learning and inline
  holography to estimate the droplet size distribution},}\ }\href@noop {}
  {\bibfield  {journal} {\bibinfo  {journal} {Int. J. Multiphase Flow}\
  }\textbf {\bibinfo {volume} {177}},\ \bibinfo {pages} {104853} (\bibinfo
  {year} {2024}{\natexlab{a}})}\BibitemShut {NoStop}%
\bibitem [{\citenamefont {Falk}\ \emph {et~al.}(2019)\citenamefont {Falk},
  \citenamefont {Mai}, \citenamefont {Bensch}, \citenamefont
  {{\c{C}}i{\c{c}}ek}, \citenamefont {Abdulkadir}, \citenamefont {Marrakchi},
  \citenamefont {B{\"{o}}hm}, \citenamefont {Deubner}, \citenamefont
  {J{\"{a}}ckel}, \citenamefont {Seiwald}, \citenamefont {Dovzhenko},
  \citenamefont {Tietz}, \citenamefont {{Dal Bosco}}, \citenamefont {Walsh},
  \citenamefont {Saltukoglu}, \citenamefont {Tay}, \citenamefont {Prinz},
  \citenamefont {Palme}, \citenamefont {Simons}, \citenamefont {Diester},
  \citenamefont {Brox},\ and\ \citenamefont {Ronneberger}}]{Falk2019}%
  \BibitemOpen
  \bibfield  {author} {\bibinfo {author} {\bibfnamefont {T.}~\bibnamefont
  {Falk}}, \bibinfo {author} {\bibfnamefont {D.}~\bibnamefont {Mai}}, \bibinfo
  {author} {\bibfnamefont {R.}~\bibnamefont {Bensch}}, \bibinfo {author}
  {\bibfnamefont {{\"{O}}.}~\bibnamefont {{\c{C}}i{\c{c}}ek}}, \bibinfo
  {author} {\bibfnamefont {A.}~\bibnamefont {Abdulkadir}}, \bibinfo {author}
  {\bibfnamefont {Y.}~\bibnamefont {Marrakchi}}, \bibinfo {author}
  {\bibfnamefont {A.}~\bibnamefont {B{\"{o}}hm}}, \bibinfo {author}
  {\bibfnamefont {J.}~\bibnamefont {Deubner}}, \bibinfo {author} {\bibfnamefont
  {Z.}~\bibnamefont {J{\"{a}}ckel}}, \bibinfo {author} {\bibfnamefont
  {K.}~\bibnamefont {Seiwald}}, \bibinfo {author} {\bibfnamefont
  {A.}~\bibnamefont {Dovzhenko}}, \bibinfo {author} {\bibfnamefont
  {O.}~\bibnamefont {Tietz}}, \bibinfo {author} {\bibfnamefont
  {C.}~\bibnamefont {{Dal Bosco}}}, \bibinfo {author} {\bibfnamefont
  {S.}~\bibnamefont {Walsh}}, \bibinfo {author} {\bibfnamefont
  {D.}~\bibnamefont {Saltukoglu}}, \bibinfo {author} {\bibfnamefont {T.~L.}\
  \bibnamefont {Tay}}, \bibinfo {author} {\bibfnamefont {M.}~\bibnamefont
  {Prinz}}, \bibinfo {author} {\bibfnamefont {K.}~\bibnamefont {Palme}},
  \bibinfo {author} {\bibfnamefont {M.}~\bibnamefont {Simons}}, \bibinfo
  {author} {\bibfnamefont {I.}~\bibnamefont {Diester}}, \bibinfo {author}
  {\bibfnamefont {T.}~\bibnamefont {Brox}}, \ and\ \bibinfo {author}
  {\bibfnamefont {O.}~\bibnamefont {Ronneberger}},\ }\bibfield  {title}
  {\enquote {\bibinfo {title} {{U-Net: deep learning for cell counting,
  detection, and morphometry}},}\ }\href {\doibase 10.1038/s41592-018-0261-2}
  {\bibfield  {journal} {\bibinfo  {journal} {Nat. Methods.}\ }\textbf
  {\bibinfo {volume} {16}},\ \bibinfo {pages} {67--70} (\bibinfo {year}
  {2019})}\BibitemShut {NoStop}%
\bibitem [{\citenamefont {Ade}\ \emph {et~al.}(2024{\natexlab{b}})\citenamefont
  {Ade}, \citenamefont {Kirar}, \citenamefont {Chandrala},\ and\ \citenamefont
  {Sahu}}]{ade2024prf}%
  \BibitemOpen
  \bibfield  {author} {\bibinfo {author} {\bibfnamefont {S.~S.}\ \bibnamefont
  {Ade}}, \bibinfo {author} {\bibfnamefont {P.~K.}\ \bibnamefont {Kirar}},
  \bibinfo {author} {\bibfnamefont {L.~D.}\ \bibnamefont {Chandrala}}, \ and\
  \bibinfo {author} {\bibfnamefont {K.~C.}\ \bibnamefont {Sahu}},\ }\bibfield
  {title} {\enquote {\bibinfo {title} {Droplet breakup and size distribution in
  an airstream: Effect of inertia},}\ }\href@noop {} {\bibfield  {journal}
  {\bibinfo  {journal} {Phys. Rev. Fluids}\ }\textbf {\bibinfo {volume} {9}},\
  \bibinfo {pages} {084004} (\bibinfo {year} {2024}{\natexlab{b}})}\BibitemShut
  {NoStop}%
\bibitem [{\citenamefont {Ade}\ \emph {et~al.}(2023)\citenamefont {Ade},
  \citenamefont {Kirar}, \citenamefont {Chandrala},\ and\ \citenamefont
  {Sahu}}]{ade2022droplet}%
  \BibitemOpen
  \bibfield  {author} {\bibinfo {author} {\bibfnamefont {S.~S.}\ \bibnamefont
  {Ade}}, \bibinfo {author} {\bibfnamefont {P.~K.}\ \bibnamefont {Kirar}},
  \bibinfo {author} {\bibfnamefont {L.~D.}\ \bibnamefont {Chandrala}}, \ and\
  \bibinfo {author} {\bibfnamefont {K.~C.}\ \bibnamefont {Sahu}},\ }\bibfield
  {title} {\enquote {\bibinfo {title} {Droplet size distribution in a swirl
  airstream using in-line holography technique},}\ }\href@noop {} {\bibfield
  {journal} {\bibinfo  {journal} {J. Fluid Mech.}\ }\textbf {\bibinfo {volume}
  {954}},\ \bibinfo {pages} {A39} (\bibinfo {year} {2023})}\BibitemShut
  {NoStop}%
\bibitem [{\citenamefont {Nicholls}\ and\ \citenamefont
  {Ranger}(1969)}]{nicholls1969aerodynamic}%
  \BibitemOpen
  \bibfield  {author} {\bibinfo {author} {\bibfnamefont {J.~A.}\ \bibnamefont
  {Nicholls}}\ and\ \bibinfo {author} {\bibfnamefont {A.~A.}\ \bibnamefont
  {Ranger}},\ }\bibfield  {title} {\enquote {\bibinfo {title} {Aerodynamic
  shattering of liquid drops.}}\ }\href@noop {} {\bibfield  {journal} {\bibinfo
   {journal} {AIAA J.}\ }\textbf {\bibinfo {volume} {7}},\ \bibinfo {pages}
  {285--290} (\bibinfo {year} {1969})}\BibitemShut {NoStop}%
\bibitem [{\citenamefont {Culick}(1960)}]{culick1960comments}%
  \BibitemOpen
  \bibfield  {author} {\bibinfo {author} {\bibfnamefont {F.~E.~C.}\
  \bibnamefont {Culick}},\ }\bibfield  {title} {\enquote {\bibinfo {title}
  {Comments on a ruptured soap film},}\ }\href@noop {} {\bibfield  {journal}
  {\bibinfo  {journal} {J. Appl. Phys.}\ }\textbf {\bibinfo {volume} {31}},\
  \bibinfo {pages} {1128--1129} (\bibinfo {year} {1960})}\BibitemShut {NoStop}%
\bibitem [{\citenamefont {Wang}\ \emph {et~al.}(2018)\citenamefont {Wang},
  \citenamefont {Dandekar}, \citenamefont {Bustos}, \citenamefont {Poulain},\
  and\ \citenamefont {Bourouiba}}]{wang2018universal}%
  \BibitemOpen
  \bibfield  {author} {\bibinfo {author} {\bibfnamefont {Y.}~\bibnamefont
  {Wang}}, \bibinfo {author} {\bibfnamefont {R.}~\bibnamefont {Dandekar}},
  \bibinfo {author} {\bibfnamefont {N.}~\bibnamefont {Bustos}}, \bibinfo
  {author} {\bibfnamefont {S.}~\bibnamefont {Poulain}}, \ and\ \bibinfo
  {author} {\bibfnamefont {L.}~\bibnamefont {Bourouiba}},\ }\bibfield  {title}
  {\enquote {\bibinfo {title} {Universal rim thickness in unsteady sheet
  fragmentation},}\ }\href@noop {} {\bibfield  {journal} {\bibinfo  {journal}
  {Phys. Rev. Lett.}\ }\textbf {\bibinfo {volume} {120}},\ \bibinfo {pages}
  {204503} (\bibinfo {year} {2018})}\BibitemShut {NoStop}%
\bibitem [{\citenamefont {Keshavarz}\ \emph {et~al.}(2020)\citenamefont
  {Keshavarz}, \citenamefont {Houze}, \citenamefont {Moore}, \citenamefont
  {Koerner},\ and\ \citenamefont {McKinley}}]{keshavarz2020rotary}%
  \BibitemOpen
  \bibfield  {author} {\bibinfo {author} {\bibfnamefont {B.}~\bibnamefont
  {Keshavarz}}, \bibinfo {author} {\bibfnamefont {E.~C.}\ \bibnamefont
  {Houze}}, \bibinfo {author} {\bibfnamefont {J.~R.}\ \bibnamefont {Moore}},
  \bibinfo {author} {\bibfnamefont {M.~R.}\ \bibnamefont {Koerner}}, \ and\
  \bibinfo {author} {\bibfnamefont {G.~H.}\ \bibnamefont {McKinley}},\
  }\bibfield  {title} {\enquote {\bibinfo {title} {Rotary atomization of
  newtonian and viscoelastic liquids},}\ }\href@noop {} {\bibfield  {journal}
  {\bibinfo  {journal} {Phys. Rev. Fluids}\ }\textbf {\bibinfo {volume} {5}},\
  \bibinfo {pages} {033601} (\bibinfo {year} {2020})}\BibitemShut {NoStop}%
\bibitem [{\citenamefont {Zhao}\ \emph {et~al.}(2010)\citenamefont {Zhao},
  \citenamefont {Liu}, \citenamefont {Li},\ and\ \citenamefont
  {Xu}}]{zhao2010morphological}%
  \BibitemOpen
  \bibfield  {author} {\bibinfo {author} {\bibfnamefont {H.}~\bibnamefont
  {Zhao}}, \bibinfo {author} {\bibfnamefont {H.~F.}\ \bibnamefont {Liu}},
  \bibinfo {author} {\bibfnamefont {W.~F.}\ \bibnamefont {Li}}, \ and\ \bibinfo
  {author} {\bibfnamefont {J.~L.}\ \bibnamefont {Xu}},\ }\bibfield  {title}
  {\enquote {\bibinfo {title} {Morphological classification of low viscosity
  drop bag breakup in a continuous air jet stream},}\ }\href@noop {} {\bibfield
   {journal} {\bibinfo  {journal} {Phys. Fluids}\ }\textbf {\bibinfo {volume}
  {22}},\ \bibinfo {pages} {114103} (\bibinfo {year} {2010})}\BibitemShut
  {NoStop}%
\bibitem [{\citenamefont {Prabhakaran}\ \emph {et~al.}(2023)\citenamefont
  {Prabhakaran}, \citenamefont {Murugavel}, \citenamefont {Konwar},
  \citenamefont {Malap}, \citenamefont {Gayatri}, \citenamefont {Dixit},
  \citenamefont {Samanta}, \citenamefont {Chowdhuri}, \citenamefont {Bera},
  \citenamefont {Varghese}, \citenamefont {Rao}, \citenamefont {Sandeep},
  \citenamefont {Safai}, \citenamefont {Sahai}, \citenamefont {Axisan},
  \citenamefont {Karipot}, \citenamefont {Baumgardner}, \citenamefont {Werden},
  \citenamefont {Fortner}, \citenamefont {Hibert}, \citenamefont {Nair},
  \citenamefont {Bankar}, \citenamefont {Gurnule}, \citenamefont {Todekar},
  \citenamefont {Jose}, \citenamefont {Jayachandran}, \citenamefont {Soyam},
  \citenamefont {Gupta}, \citenamefont {Choudhary}, \citenamefont
  {Aravindhavel}, \citenamefont {Kantipudi}, \citenamefont {Pradeepkumar},
  \citenamefont {Krishnan}, \citenamefont {Nandakumar}, \citenamefont
  {DeCarlo}, \citenamefont {Worsnop}, \citenamefont {Bhat}, \citenamefont
  {Rajeevan},\ and\ \citenamefont {Nanjundiah}}]{prabhakaran2023caipeex}%
  \BibitemOpen
  \bibfield  {author} {\bibinfo {author} {\bibfnamefont {T.}~\bibnamefont
  {Prabhakaran}}, \bibinfo {author} {\bibfnamefont {P.}~\bibnamefont
  {Murugavel}}, \bibinfo {author} {\bibfnamefont {M.}~\bibnamefont {Konwar}},
  \bibinfo {author} {\bibfnamefont {N.}~\bibnamefont {Malap}}, \bibinfo
  {author} {\bibfnamefont {K.}~\bibnamefont {Gayatri}}, \bibinfo {author}
  {\bibfnamefont {S.}~\bibnamefont {Dixit}}, \bibinfo {author} {\bibfnamefont
  {S.}~\bibnamefont {Samanta}}, \bibinfo {author} {\bibfnamefont
  {S.}~\bibnamefont {Chowdhuri}}, \bibinfo {author} {\bibfnamefont
  {S.}~\bibnamefont {Bera}}, \bibinfo {author} {\bibfnamefont {M.}~\bibnamefont
  {Varghese}}, \bibinfo {author} {\bibfnamefont {J.}~\bibnamefont {Rao}},
  \bibinfo {author} {\bibfnamefont {J.}~\bibnamefont {Sandeep}}, \bibinfo
  {author} {\bibfnamefont {P.~D.}\ \bibnamefont {Safai}}, \bibinfo {author}
  {\bibfnamefont {A.~K.}\ \bibnamefont {Sahai}}, \bibinfo {author}
  {\bibfnamefont {D.}~\bibnamefont {Axisan}}, \bibinfo {author} {\bibfnamefont
  {A.}~\bibnamefont {Karipot}}, \bibinfo {author} {\bibfnamefont
  {D.}~\bibnamefont {Baumgardner}}, \bibinfo {author} {\bibfnamefont
  {B.}~\bibnamefont {Werden}}, \bibinfo {author} {\bibfnamefont
  {E.}~\bibnamefont {Fortner}}, \bibinfo {author} {\bibfnamefont
  {K.}~\bibnamefont {Hibert}}, \bibinfo {author} {\bibfnamefont
  {S.}~\bibnamefont {Nair}}, \bibinfo {author} {\bibfnamefont {S.}~\bibnamefont
  {Bankar}}, \bibinfo {author} {\bibfnamefont {D.}~\bibnamefont {Gurnule}},
  \bibinfo {author} {\bibfnamefont {K.}~\bibnamefont {Todekar}}, \bibinfo
  {author} {\bibfnamefont {J.}~\bibnamefont {Jose}}, \bibinfo {author}
  {\bibfnamefont {V.}~\bibnamefont {Jayachandran}}, \bibinfo {author}
  {\bibfnamefont {P.~S.}\ \bibnamefont {Soyam}}, \bibinfo {author}
  {\bibfnamefont {A.}~\bibnamefont {Gupta}}, \bibinfo {author} {\bibfnamefont
  {H.}~\bibnamefont {Choudhary}}, \bibinfo {author} {\bibnamefont
  {Aravindhavel}}, \bibinfo {author} {\bibfnamefont {S.~B.}\ \bibnamefont
  {Kantipudi}}, \bibinfo {author} {\bibfnamefont {P.}~\bibnamefont
  {Pradeepkumar}}, \bibinfo {author} {\bibfnamefont {R.}~\bibnamefont
  {Krishnan}}, \bibinfo {author} {\bibfnamefont {K.}~\bibnamefont
  {Nandakumar}}, \bibinfo {author} {\bibfnamefont {P.~F.}\ \bibnamefont
  {DeCarlo}}, \bibinfo {author} {\bibfnamefont {D.}~\bibnamefont {Worsnop}},
  \bibinfo {author} {\bibfnamefont {G.~S.}\ \bibnamefont {Bhat}}, \bibinfo
  {author} {\bibfnamefont {M.}~\bibnamefont {Rajeevan}}, \ and\ \bibinfo
  {author} {\bibfnamefont {R.}~\bibnamefont {Nanjundiah}},\ }\bibfield  {title}
  {\enquote {\bibinfo {title} {{CAIPEEX:} indian cloud seeding scientific
  experiment},}\ }\href@noop {} {\bibfield  {journal} {\bibinfo  {journal}
  {Bull. Amer. Meteor. Soc.}\ }\textbf {\bibinfo {volume} {104}},\ \bibinfo
  {pages} {E2095--E2120} (\bibinfo {year} {2023})}\BibitemShut {NoStop}%
\bibitem [{\citenamefont {Patade}\ \emph {et~al.}(2015)\citenamefont {Patade},
  \citenamefont {Prabha}, \citenamefont {Axisa}, \citenamefont {Gayatri},\ and\
  \citenamefont {Heymsfield}}]{patade2015particle}%
  \BibitemOpen
  \bibfield  {author} {\bibinfo {author} {\bibfnamefont {S.}~\bibnamefont
  {Patade}}, \bibinfo {author} {\bibfnamefont {T.~V.}\ \bibnamefont {Prabha}},
  \bibinfo {author} {\bibfnamefont {D.}~\bibnamefont {Axisa}}, \bibinfo
  {author} {\bibfnamefont {K.}~\bibnamefont {Gayatri}}, \ and\ \bibinfo
  {author} {\bibfnamefont {A.}~\bibnamefont {Heymsfield}},\ }\bibfield  {title}
  {\enquote {\bibinfo {title} {Particle size distribution properties in
  mixed-phase monsoon clouds from in situ measurements during {CAIPEEX}},}\
  }\href@noop {} {\bibfield  {journal} {\bibinfo  {journal} {J. Geophys. Res.
  Atmos.}\ }\textbf {\bibinfo {volume} {120}},\ \bibinfo {pages} {10--418}
  (\bibinfo {year} {2015})}\BibitemShut {NoStop}%
\bibitem [{\citenamefont {Dye}\ and\ \citenamefont
  {Baumgardner}(1984)}]{dye1984evaluation}%
  \BibitemOpen
  \bibfield  {author} {\bibinfo {author} {\bibfnamefont {J.~E.}\ \bibnamefont
  {Dye}}\ and\ \bibinfo {author} {\bibfnamefont {D.}~\bibnamefont
  {Baumgardner}},\ }\bibfield  {title} {\enquote {\bibinfo {title} {Evaluation
  of the forward scattering spectrometer probe. part i: Electronic and optical
  studies},}\ }\href@noop {} {\bibfield  {journal} {\bibinfo  {journal} {J.
  Atmos. Ocean. Technol.}\ }\textbf {\bibinfo {volume} {1}},\ \bibinfo {pages}
  {329--344} (\bibinfo {year} {1984})}\BibitemShut {NoStop}%
\bibitem [{\citenamefont {Baumgardner}\ \emph {et~al.}(2001)\citenamefont
  {Baumgardner}, \citenamefont {Jonsson}, \citenamefont {Dawson}, \citenamefont
  {O'Connor},\ and\ \citenamefont {Newton}}]{baumgardner2001cloud}%
  \BibitemOpen
  \bibfield  {author} {\bibinfo {author} {\bibfnamefont {D.}~\bibnamefont
  {Baumgardner}}, \bibinfo {author} {\bibfnamefont {H.}~\bibnamefont
  {Jonsson}}, \bibinfo {author} {\bibfnamefont {W.}~\bibnamefont {Dawson}},
  \bibinfo {author} {\bibfnamefont {D.}~\bibnamefont {O'Connor}}, \ and\
  \bibinfo {author} {\bibfnamefont {R.}~\bibnamefont {Newton}},\ }\bibfield
  {title} {\enquote {\bibinfo {title} {The cloud, aerosol and precipitation
  spectrometer: a new instrument for cloud investigations},}\ }\href@noop {}
  {\bibfield  {journal} {\bibinfo  {journal} {Atmos. Res.}\ }\textbf {\bibinfo
  {volume} {59}},\ \bibinfo {pages} {251--264} (\bibinfo {year}
  {2001})}\BibitemShut {NoStop}%
\bibitem [{\citenamefont {Patade}\ \emph {et~al.}(2016)\citenamefont {Patade},
  \citenamefont {Shete}, \citenamefont {Malap}, \citenamefont {Kulkarni},\ and\
  \citenamefont {Prabha}}]{patade2016observational}%
  \BibitemOpen
  \bibfield  {author} {\bibinfo {author} {\bibfnamefont {S.}~\bibnamefont
  {Patade}}, \bibinfo {author} {\bibfnamefont {S.}~\bibnamefont {Shete}},
  \bibinfo {author} {\bibfnamefont {N.}~\bibnamefont {Malap}}, \bibinfo
  {author} {\bibfnamefont {G.}~\bibnamefont {Kulkarni}}, \ and\ \bibinfo
  {author} {\bibfnamefont {T.}~\bibnamefont {Prabha}},\ }\bibfield  {title}
  {\enquote {\bibinfo {title} {Observational and simulated cloud microphysical
  features of rain formation in the mixed phase clouds observed during
  caipeex},}\ }\href@noop {} {\bibfield  {journal} {\bibinfo  {journal}
  {Atmospheric Research}\ }\textbf {\bibinfo {volume} {169}},\ \bibinfo {pages}
  {32--45} (\bibinfo {year} {2016})}\BibitemShut {NoStop}%
\bibitem [{\citenamefont {Raut}\ \emph {et~al.}(2021)\citenamefont {Raut},
  \citenamefont {Konwar}, \citenamefont {Murugavel}, \citenamefont {Kadge},
  \citenamefont {Gurnule}, \citenamefont {Sayyed}, \citenamefont {Todekar},
  \citenamefont {Malap}, \citenamefont {Bankar},\ and\ \citenamefont
  {Prabhakaran}}]{raut2021microphysical}%
  \BibitemOpen
  \bibfield  {author} {\bibinfo {author} {\bibfnamefont {B.~A.}\ \bibnamefont
  {Raut}}, \bibinfo {author} {\bibfnamefont {M.}~\bibnamefont {Konwar}},
  \bibinfo {author} {\bibfnamefont {P.}~\bibnamefont {Murugavel}}, \bibinfo
  {author} {\bibfnamefont {D.}~\bibnamefont {Kadge}}, \bibinfo {author}
  {\bibfnamefont {D.}~\bibnamefont {Gurnule}}, \bibinfo {author} {\bibfnamefont
  {I.}~\bibnamefont {Sayyed}}, \bibinfo {author} {\bibfnamefont
  {K.}~\bibnamefont {Todekar}}, \bibinfo {author} {\bibfnamefont
  {N.}~\bibnamefont {Malap}}, \bibinfo {author} {\bibfnamefont
  {S.}~\bibnamefont {Bankar}}, \ and\ \bibinfo {author} {\bibfnamefont
  {T.}~\bibnamefont {Prabhakaran}},\ }\bibfield  {title} {\enquote {\bibinfo
  {title} {Microphysical origin of raindrop size distributions during the
  indian monsoon},}\ }\href@noop {} {\bibfield  {journal} {\bibinfo  {journal}
  {Geophys. Res. Lett.}\ }\textbf {\bibinfo {volume} {48}},\ \bibinfo {pages}
  {e2021GL093581} (\bibinfo {year} {2021})}\BibitemShut {NoStop}%
\bibitem [{\citenamefont {Srivastava}(1978)}]{srivastava1978parameterization}%
  \BibitemOpen
  \bibfield  {author} {\bibinfo {author} {\bibfnamefont {R.~C.}\ \bibnamefont
  {Srivastava}},\ }\bibfield  {title} {\enquote {\bibinfo {title}
  {Parameterization of raindrop size distributions},}\ }\href@noop {}
  {\bibfield  {journal} {\bibinfo  {journal} {J. Atmos. Sci.}\ }\textbf
  {\bibinfo {volume} {35}},\ \bibinfo {pages} {108--117} (\bibinfo {year}
  {1978})}\BibitemShut {NoStop}%
\bibitem [{\citenamefont {Low}\ and\ \citenamefont
  {List}(1982)}]{low1982collision}%
  \BibitemOpen
  \bibfield  {author} {\bibinfo {author} {\bibfnamefont {T.~B.}\ \bibnamefont
  {Low}}\ and\ \bibinfo {author} {\bibfnamefont {R.}~\bibnamefont {List}},\
  }\bibfield  {title} {\enquote {\bibinfo {title} {Collision, coalescence and
  breakup of raindrops. part ii: Parameterization of fragment size
  distributions},}\ }\href@noop {} {\bibfield  {journal} {\bibinfo  {journal}
  {J. Atmos. Sci.}\ }\textbf {\bibinfo {volume} {39}},\ \bibinfo {pages}
  {1607--1619} (\bibinfo {year} {1982})}\BibitemShut {NoStop}%
\bibitem [{\citenamefont {McFarquhar}(2004)}]{mcfarquhar2004new}%
  \BibitemOpen
  \bibfield  {author} {\bibinfo {author} {\bibfnamefont {G.~M.}\ \bibnamefont
  {McFarquhar}},\ }\bibfield  {title} {\enquote {\bibinfo {title} {A new
  representation of collision-induced breakup of raindrops and its implications
  for the shapes of raindrop size distributions},}\ }\href@noop {} {\bibfield
  {journal} {\bibinfo  {journal} {J. Atmos. Sci.}\ }\textbf {\bibinfo {volume}
  {61}},\ \bibinfo {pages} {777--794} (\bibinfo {year} {2004})}\BibitemShut
  {NoStop}%
\bibitem [{\citenamefont {Straub}\ \emph {et~al.}(2010)\citenamefont {Straub},
  \citenamefont {Beheng}, \citenamefont {Seifert}, \citenamefont {Schlottke},\
  and\ \citenamefont {Weigand}}]{straub2010numerical}%
  \BibitemOpen
  \bibfield  {author} {\bibinfo {author} {\bibfnamefont {W.}~\bibnamefont
  {Straub}}, \bibinfo {author} {\bibfnamefont {K.~D.}\ \bibnamefont {Beheng}},
  \bibinfo {author} {\bibfnamefont {A.}~\bibnamefont {Seifert}}, \bibinfo
  {author} {\bibfnamefont {J.}~\bibnamefont {Schlottke}}, \ and\ \bibinfo
  {author} {\bibfnamefont {B.}~\bibnamefont {Weigand}},\ }\bibfield  {title}
  {\enquote {\bibinfo {title} {Numerical investigation of collision-induced
  breakup of raindrops. part ii: Parameterizations of coalescence efficiencies
  and fragment size distributions},}\ }\href@noop {} {\bibfield  {journal}
  {\bibinfo  {journal} {J. Atmos. Sci.}\ }\textbf {\bibinfo {volume} {67}},\
  \bibinfo {pages} {576--588} (\bibinfo {year} {2010})}\BibitemShut {NoStop}%
\bibitem [{\citenamefont {Testik}\ and\ \citenamefont
  {Rahman}(2017)}]{testik2017first}%
  \BibitemOpen
  \bibfield  {author} {\bibinfo {author} {\bibfnamefont {F.~Y.}\ \bibnamefont
  {Testik}}\ and\ \bibinfo {author} {\bibfnamefont {M.~K.}\ \bibnamefont
  {Rahman}},\ }\bibfield  {title} {\enquote {\bibinfo {title} {First in situ
  observations of binary raindrop collisions},}\ }\href@noop {} {\bibfield
  {journal} {\bibinfo  {journal} {Geophys. Res. Lett.}\ }\textbf {\bibinfo
  {volume} {44}},\ \bibinfo {pages} {1175--1181} (\bibinfo {year}
  {2017})}\BibitemShut {NoStop}%
\end{thebibliography}

%

\end{document}